\documentclass[final,1p,sort&compress,times]{elsarticle}

\usepackage{amssymb}
\usepackage{hyperref}

\journal{Physics Letters A}

\begin{document}

\begin{frontmatter}



\title{Convergence of macrostates under reproducible processes}


\author{Jochen Rau}
\ead{jochen.rau@q-info.org}
\ead[url]{www.q-info.org}

\address{Institut f\"ur Theoretische Physik,
Johann Wolfgang Goethe-Universit\"at,
Max-von-Laue-Str. 1, 
60438 Frankfurt am Main,
Germany}

\begin{abstract}
I show that whenever a  system undergoes a reproducible macroscopic process the mutual distinguishability of macrostates, as measured by their relative entropy, diminishes.
This extends the second law which regards only ordinary entropies, and hence only the distinguishability between macrostates and one specific reference state (equidistribution).
The new result holds regardless of whether the  process is linear or nonlinear.
Its proof hinges on the monotonicity of quantum relative entropy under arbitrary coarse grainings, even those that cannot be represented by completely positive maps.

\end{abstract}


\end{frontmatter}



\section{\label{intro}Motivation}

As discussed lucidly by Jaynes \cite{jaynes:2ndlaw} the second law of thermodynamics follows directly from the requirement that macroscopic processes be reproducible.
The microscopic details of a macroscopic system are generally beyond the reach of an experimenter;
all she can observe and control is some limited set of macroscopic variables, typically pertaining to the constants of the motion or, if out of equilibrium, slowly varying degrees of freedom \cite{rau:physrep}.
The system is  modelled by a (generalized) canonical distribution 
\begin{equation}
	\mu_{g} \propto \exp \left[\sum_a \lambda^a G_a\right]
	,
\label{canonical}
\end{equation}
where $\{G_a\}$ are the observables whose expectation values  characterize the system's macrostate. 
With properly adjusted Lagrange parameters $\{\lambda^a\}$ 
this canonical state encodes information about the ``relevant'' expectation values $\{g_a\equiv \langle G_a\rangle_\mu\}$,
while discarding (by maximising entropy) all other information.
The preparation of the system  amounts to setting the initial values of the relevant expectation values or of their conjugate Lagrange parameters.
The system then undergoes a macroscopic process, at the end of which its macrostate is again characterized by the expectation values pertaining to some small (possibly different) set of observables $\{{F}_b\}$. 
The process is reproducible if the  preparation uniquely determines the outcome;
i.e., if merely on the basis of the initial $\{g_a\}$ one can predict the final $\{{f}_b\}$.
A prediction cannot possibly contain more information than the data on which it is based.
So the final expectation values cannot carry more information than do the initial expectation values;
and hence indeed, the corresponding final entropy cannot be smaller than the initial entropy.

The above argument translates into a simple mathematical proof of the second law as follows.
By assumption the initial state is modelled by $\mu_g$.
On the microscopic level (and possibly enlarging the system to include any environment it might be coupled to) every process is unitary,
$\mu_g\to U\mu_g U^\dagger$, preserving entropy: $S[U\mu_g U^\dagger]=S[\mu_g]$.
Again by assumption, the final state's macroscopic features  can be described just as well by the canonical distribution $\mu_{{f}}$, where the relevant expectation values must be consistent with the underlying unitary dynamics:
\begin{equation}
	{f}_b=\langle F_b\rangle_{U\mu_g U^\dagger}
	.
\end{equation}
The replacement $U\mu_g U^\dagger \to \mu_f$ constitutes a coarse graining that discards information about all degrees of freedom but the relevant expectation values;
it can thus only increase the entropy, $S[\mu_f]\geq S[U\mu_g U^\dagger]$.
Hence $S[\mu_f]\geq S[\mu_g]$, Q.E.D.

Rather than in terms of  ordinary entropy the second law can also be formulated in terms of  \textit{relative} entropy.
The relative entropy between two states $\rho$ and $\sigma$,
\begin{equation}
	S(\rho\|\sigma):=\mbox{tr}(\rho\ln\rho-\rho\ln\sigma)
	,
\end{equation}
is a measure for their distinguishability  \cite{vedral:rmp}.
A precise operational interpretation is furnished by the {quantum Stein lemma} \cite{hiai+petz,ogawa+nagaoka,brandao+plenio}:
The probability that, given a sample of size $N$ taken from an i.i.d. source of states $\sigma$,
tomography will erroneously reveal the state $\rho$ decreases exponentially with $N$ as $N\to\infty$,
\begin{equation}
	\mbox{prob}_\epsilon(\rho^{ N}|\sigma^{ N})\sim \exp[-N S(\rho\|\sigma)]
\end{equation}
regardless of the  threshold $\epsilon$ ($0<\epsilon<1$),
where 
\begin{equation}
	\mbox{prob}_\epsilon(\rho^{ N}|\sigma^{ N}):=
	\inf_{0\leq\Gamma\leq 1} \left.\left\{(\sigma^{ N}|\Gamma)\right|(\rho^{ N}|\Gamma)\geq \epsilon\right\}
	,
\label{defprob}
\end{equation}
$(X|Y):=\mbox{tr}(X^\dagger Y)$
and, e.g., $\rho^N$ is short for $\rho^{\otimes N}$.
Loosely speaking, therefore, the relevant entropy drives  the probability of mistaking $\sigma$ for $\rho$.
With this definition of distinguishability the statement of the second law -- that any reproducible process increases entropy  --
is now equivalent to saying that under a reproducible process macrostates become less distinguishable  from the uniform distribution:
\begin{equation}
	S(\mu_f\|1/\mbox{tr}1) \leq S(\mu_{{g}}\|1/\mbox{tr}1)
	.
\end{equation}

As the second law reflects the tendency of macrostates to move closer to equidistribution, one would intuitively expect that
they have a tendency to converge, and  that hence not only  
the distinguishability between any macrostate and the uniform distribution (which is one particular macrostate) diminishes
but also the {mutual} distinguishability between {\textit{arbitrary} pairs} of macrostates.
Such would mean that for any two initial macrostates $\mu_g$ and $\mu_{g'}$ evolving under the same reproducible process to final macrostates $\mu_f$ and $\mu_{f'}$, respectively,
the relative entropy can only decrease:
\begin{equation}
	S(\mu_f\|\mu_{f'}) \leq S(\mu_g\|\mu_{g'})
	.
\label{2ndlawextended}
\end{equation}
This assertion is not implied by the second law;
rather, if true, it would constitute a nontrivial extension.
In fact, it is  known to be true for macroscopic processes that are linear:
A linear process is represented by a completely positive map $\Phi$ for which by Lindblad's theorem \cite{lindblad:monotonicity} 
\begin{equation}
	S(\Phi(\rho)\|\Phi(\sigma))\leq S(\rho\|\sigma)
\end{equation}
for arbitrary $\rho$ and $\sigma$.
Yet not all macroscopic processes are linear --- many real processes such as, say, the dynamics of a fluid are governed by transport equations that are nonlinear.
For this more general situation I am not aware of a proof of the inequality (\ref{2ndlawextended}).
Closing this gap is the purpose of my Letter.

\section{Derivation}

My line of argument is inspired by the above simple proof of the second law.
Like the ordinary entropy, relative entropy is invariant under unitary microscopic evolution, so
\begin{equation}
	S(U\mu_g U^\dagger\|U\mu_{g'}U^\dagger) = S(\mu_g\|\mu_{g'})
	.
\end{equation}
By assumption, for the purposes of a macroscopic description the transformed states on the left hand side may be replaced by $\mu_f$ and $\mu_{f'}$, respectively,
where the coarse grainings $U\mu_g U^\dagger\to\mu_{f}$ and $U\mu_{g'}U^\dagger\to\mu_{f'}$  preserve the respective expectation values of the relevant observables $\{F_b\}$.
Then \`a la Jaynes the inequality (\ref{2ndlawextended})  follows immediately if only one can show 
\begin{equation}
	S(\mu_f\|\mu_{f'}) \leq S(U\mu_g U^\dagger\|U\mu_{g'}U^\dagger)
	;
\end{equation}
or more generally, 
\begin{equation}
	S(\mu_{f(\rho)}\|\mu_{f(\sigma)})\leq S(\rho\|\sigma)
\label{monotonicity}
\end{equation}
with $\{f_b(\rho)\equiv \langle F_b\rangle_\rho\}$ and $\{f_b(\sigma)\equiv \langle F_b\rangle_\sigma\}$.
Such monotonicity appears  plausible because
one  intuitively expects coarse graining to diminish the distinguishability of quantum states;
and indeed, it is known to be true (again by Lindblad's theorem) whenever the coarse graining can be represented by a completely positive map.
Yet again, not all coarse grainings are of this form.
For example, if the system under consideration is bipartite ($A\times B$)  and the relevant observables are all those pertaining to 
\textit{either} $A$ or $B$ then coarse graining amounts to removing  correlations, $\rho_{AB}\to\rho_A\times \rho_B$, which is not a linear map.
The remainder of this Section shall be devoted to showing that monotonicity holds in these nonlinear cases, too.

The proof will make use of the existence and properties of the so-called {Kawasaki-Gunton projector}, 
a mathematical object originally introduced and used in the context of nonlinear transport equations \cite{rau:physrep,kawasaki+gunton}.
A special case of such a Kawasaki-Gunton projector is 
the (super-)projector ${\cal P}^{(N)}_{f(\rho)}$ which acts on the space of observables such that for arbitrary positive $\Gamma$ ($0\leq\Gamma\leq 1$)
\begin{equation}
	(\rho^{ N}|{\cal P}^{(N)}_{f(\rho)}\Gamma)=(\mu_{f(\rho)}^{ N}|\Gamma)
	.
\end{equation}
This projector  has a nontrivial form that depends on $N$ and on the relevant expectation values.
Its main properties of interest here are that 
(i) it is always linear, even when the corresponding coarse graining in state space is not;
(ii) it preserves positivity, so $\{{\cal P}\Gamma\}\subseteq \{\Gamma\}$;
and
(iii) it is idempotent in the sense
\begin{equation}
	{\cal P}^{(N)}_{f(\rho)} {\cal P}^{(N)}_{f(\sigma)} = {\cal P}^{(N)}_{f(\sigma)}
\end{equation}
even for $\rho\neq\sigma$.

By the quantum Stein lemma, proving the monotonicity (\ref{monotonicity}) is tantamount to showing that, asymptotically,
\begin{equation}
	\mbox{prob}_{\epsilon'}(\mu_{f(\rho)}^{ N}|\mu_{f(\sigma)}^{ N})
	\geq
	\mbox{prob}_\epsilon(\rho^{ N}|\sigma^{ N})
\label{monotonicity2}
\end{equation}
for conveniently chosen $\epsilon'$ and $\epsilon$.
With the definition (\ref{defprob}) and the complement ${\cal Q}:=1-{\cal P}$ of the Kawasaki-Gunton projector the left-hand side of this inequality can be written as
\begin{equation}
	\mbox{prob}_{\epsilon'}(\mu_{f(\rho)}^{ N}|\mu_{f(\sigma)}^{ N})
	=
	\inf_{0\leq\Gamma\leq 1} \left.\left\{(\sigma^{ N}|{\cal P}^{(N)}_{f(\sigma)}\Gamma)\right|(\rho^{ N}|{\cal P}^{(N)}_{f(\sigma)}\Gamma)\geq \epsilon'
	-(\mu_{f(\rho)}^{ N}|{\cal Q}^{(N)}_{f(\sigma)}\Gamma)
	\right\}
	.
\label{probcoarse}
\end{equation}
Whenever the coarse graining is represented by a completely positive map the Kawasaki-Gunton projector becomes independent of the expectation values $\{f_b\}$.
The term $(\mu|{\cal Q}\Gamma)$ then vanishes, and for the remainder the infimum can be taken just as well over $\{{\cal P}\Gamma\}$ rather than $\{\Gamma\}$.
As $\inf_{{\cal P}\Gamma}\geq\inf_\Gamma$ this implies immediately the monotonicity (\ref{monotonicity2}).
As soon as the coarse graining is nonlinear, however, further analysis of the term $(\mu|{\cal Q}\Gamma)$ is needed, to which I turn next.

For any  state $\rho$ and any canonical state $\mu_f$ I define 
\begin{equation}
	\gamma(\rho,\mu_f):=\sup_N \sup_{0\leq \Gamma\leq 1} |(\rho^{ N}|{\cal Q}^{(N)}_{f}\Gamma)|
	.
\end{equation}
This quantity  lies in the range $[0,1]$;
it vanishes at $\rho=\mu_f$;
and it can attain the maximum value $\gamma=1$, if at all, only for $\rho$ that have at least one zero eigenvalue.
If $\rho$ is replaced by $\mu_{f(\rho)}$ then the latter can only be the case if one of the associated expectation values $\{f_b(\rho)\}$ (which are different from the $\{f_b\}$) takes an extremal value;
so as long as one excludes extremal expectation values, $\gamma(\mu_{f(\rho)},\mu_f)$ is always strictly smaller than one:
\begin{equation}
	0\leq \gamma(\mu_{f(\rho)},\mu_f) < 1
	\quad\forall\quad
	f_b(\rho)\in (f_b^{\rm min},f_b^{\rm max})
	.
\end{equation}
Consequently, the choices
\begin{equation}
	\epsilon'=(1+\gamma(\mu_{f(\rho)},\mu_{f(\sigma)}))/2 
\end{equation}
\begin{equation}
	\epsilon=(1-\gamma(\mu_{f(\rho)},\mu_{f(\sigma)}))/2
\end{equation}
both lie in $(0,1)$ and  ensure that for any $N$ and  $\Gamma$
\begin{equation}
	\epsilon' -(\mu_{f(\rho)}^{ N}|{\cal Q}^{(N)}_{f(\sigma)}\Gamma) \geq \epsilon
	.
\end{equation}
Thus from Eq. (\ref{probcoarse}),
\begin{equation}
	\mbox{prob}_{\epsilon'}(\mu_{f(\rho)}^{ N}|\mu_{f(\sigma)}^{ N})
	\geq
	\inf_{0\leq\Gamma\leq 1} \left.\left\{(\sigma^{ N}|{\cal P}^{(N)}_{f(\sigma)}\Gamma)\right|(\rho^{ N}|{\cal P}^{(N)}_{f(\sigma)}\Gamma)\geq \epsilon\right\}
	.	
\end{equation}
On the right-hand side the infimum can now be taken just as well over $\{{\cal P}^{(N)}_{f(\sigma)}\Gamma\}$ rather than $\{\Gamma\}$;
and as $\inf_{{\cal P}\Gamma}\geq\inf_\Gamma$ this implies the monotonicity (\ref{monotonicity2}), Q.E.D.

\section{Conclusion}

Under a reproducible process macrostates have a tendency not only to move closer towards equidistribution (as stated by the second law)
but also to move closer towards each other, as I have shown in this Letter.
The proof depended crucially on the monotonicity of quantum relative entropy under arbitrary (including nonlinear) coarse grainings.
This generalized monotonicity is an interesting result in itself, with useful implications not only for statistical mechanics but also 
for quantum information theory.
For instance, in the earlier example of coarse graining by removing correlations it implies
\begin{equation}
	S(\rho_A\times \rho_B\|\sigma_A\times \sigma_B)\leq S(\rho_{AB}\|\sigma_{AB})
	,
\end{equation}
a much stronger result than the well-known monotonicity $S(\rho_A\|\sigma_A)\leq S(\rho_{AB}\|\sigma_{AB})$.


\bibliographystyle{elsarticle-num}


\end{document}